\documentclass[12pt]{iopart}

\usepackage{graphicx}
\expandafter\let\csname equation*\endcsname=\relax
\expandafter\let\csname endequation*\endcsname=\relax
\usepackage{amsmath}

\begin{document}

\title[]{Near infrared few-cycle pulses for high harmonic generation}

\author{Steffen Driever$^{a}$, Konstantin B. Holzner$^{a}$, Jean-Christophe Delagnes$^{b}$, Nikita Fedorov$^{b}$, Martin Arnold$^{a}$, Damien Bigourd$^{a}$, Eric Cormier$^{b}$, Roland Guichard$^{a}$, Eric Constant$^{b}$ and Amelle Za\"{i}r$^{a}$ 
}
\address{
$^{a}$Imperial College London, Department of Physics, Blackett Laboratory Laser Consortium, London SW7 2AZ United Kingdom}

\address{$^{b}$Universit$\acute{e}$ de Bordeaux, CEA, CNRS UMR 5107, CELIA (Centre Lasers Intenses et Applications), FR-33400 Talence, France}

\ead{steffen.driever10@imperial.ac.uk}
\begin{abstract}
We report on the development of tunable few-cycle pulses with central wavelengths from 1.6\,$\mu m$ to 2\,$\mu m$. Theses pulses were used as a proof of principle for high harmonic generation in atomic and molecular targets.  In order to generate such pulses we produced a filament in a 4 bar krypton cell. Spectral broadening by a factor of 2\,-\,3 of a 40\,fs near infrared input pulse was achieved. The spectrally broadened output pulses were then compressed by fused silica plates down to the few-cycle regime close to the Fourier limit. The auto-correlation of these pulses revealed durations of $\sim$\,3 cycles for all investigated central wavelengths. Pulses with a central wavelength of 1.7\,$\mu m$ and up to 430\,$\mu J$ energy per pulse were employed to generate high order harmonics in Xe, Ar and N$_2$. Moving to near infrared few-cycle pulses opens the possibility to operate deeply in the non-perturbative regime with a Keldysh parameter $\gamma << 1$. Hence, this source is suitable for the study of the non-adiabatic tunneling regime in most generating systems used for high order harmonic generation and attoscience.
\end{abstract}
\maketitle

\section{Introduction: }

During the interaction of intense laser fields with atoms and molecules an electron wavepacket initially bound to the target can be ionized through different interaction processes. These ionization channels determine the evolution of the core and the electron wavepacket in the continuum. To identify which of these processes is predominant during the interaction, L. V. Keldysh \cite{keldysh} introduced a parameter given by $\gamma=\sqrt{I_p/2U_p}$ where $I_p$ is the target's ionization potential and $U_p$ is the ponderomotive energy associated with the strong laser field employed.

In the perturbative regime, $\gamma>>1$, the target's potential remains largely unchanged by the laser field. In this case multi-photon processes are necessary to emit a single electron wavepacket from the ground state. The electron wavepacket can absorb many photons, reaching energies just exceeding the ionization potential (MPI: multi-photon ionization), or absorb many more photons to overcome the ionization potential. After ionization, the electron wavepacket in the continuum can be considered only subject to the laser field \cite{lewenstein94, corkum93, kulander}. When the laser field reverses half a cycle later the electron wavepacket can be redirected to the ionic core and then re-scatter (ATI: Above Threshold Ionization \cite{agostini}).

 When $\gamma<<1$, known as the non-perturbative regime, the target's potential is deformed by the strong laser field. This deformation happens at every maximum of the laser field, i.e. twice per optical cycle. Every optical half-cycle, the electron wavepacket initially bound to the target, sees the tail of the Coulomb potential evolve in time, hence creating a barrier whose height and width change over time. Thus, the electron wavepacket reaches a greater probability to tunnel out of the system through this potential barrier. The probability of tunneling through the barrier increases with the laser field strength to the point where the barrier drops below the binding energy level and the electron wavepacket is directly launched into the continuum (BSI: Barrier Suppression Ionization). The latter situation is the asymptotic limit of the tunnel ionization regime and usually results in an electron wavepacket which cannot be redirected by the laser field back to the ionic core. As a result recombination or re-scattering of the electron wavepacket does not occur.

In order to understand how ionization occurs we need to follow the transition between the different regimes of ionization.
 To study these transitions, most of the experiments were performed using an IR laser field (0.8\,$\mu m$) with a typical intensity of $I=10^{14}\,W/cm^{2}$, interacting with atoms (noble gases) \cite{corkum93, kulander,huillier92, huillier93}, molecules \cite{marangos}, nanoparticles \cite{kim,shaaran,sussmann} or periodic solid structures \cite{schiffrin,apalkov}. For these targets, the ionization potential is kept below $\sim$ 25\,eV, so that ionization regimes correspond to $\gamma$ values between 0.5 and $\sim$ 5. Under such conditions, only high $I_p$ atoms subject to a strong laser field are in the non-perturbative regime. To enable the study of dynamical processes under the non-perturbative regime for most atoms and molecules two routes can be followed.\\

Firstly, one can increase the laser field intensity keeping the wavelength in the infrared (IR) region (0.8\,$\mu m$ for instance). This approach has been employed widely owing to the development of Ti:Sa laser technologies \cite{second_tisa86}. However it is strongly limited by the field strength that a target can withstand before its ground state is strongly depleted. One way to circumvent this limitation is to use few-cycle pulses and to control their intensity profile. In this case one has to reconsider the Keldysh parameter in a non-quasi static situation as shown by Yudin and Ivanov \cite{ivanov}. They defined a general expression for non-adiabatic ionization rates based on a modification of the ADK theory \cite{adk} suitable for the case of few-cycle laser fields.\\

The second route is to decrease $\gamma$ by increasing wavelengths ($\lambda$) to the near (NIR )or mid-infrared (MIR) region whilst maintaining low enough intensity to avoid depletion of the ground state. 
As a result it is possible to study the ultrafast dynamics of charge migration involved in atoms and molecules after ionization using high harmonic generation (HHG) \cite {baker06,marangos08,lein07,itatani08,hassler10,torres10, murnane09, olga09, zair12}. However, analyzing HHG driven by few-cycle NIR to MIR pulses offers the possibility to extend these studies well into the non-perturbative regime and study how these charge migrations evolve while transitioning from one regime of ionization to another.

\section{Employing a NIR to MIR laser field for high order harmonic generation}
In recent years, efforts have been made to push the development of multi-cycle NIR to MIR femtosecond sources to drive strong field interaction such as HHG experiments. Indeed, using laser fields at longer wavelengths, i.e. in the low frequency limit to generate high order harmonics has the potential to dramatically extend the maximum photon energy produced (cut-off position). This is because it scales with $\propto\lambda^2$  \cite{lewenstein94,corkum93, krause92, lewenstein95} and goes beyond the carbon K-edge. Hence, the coherent XUV and X-ray radiation produced can provide a unique source to study ultrafast inner core dynamics \cite{drescher} opening the possibility to produce isolated attosecond pulses in the soft X-ray regime \cite{chen, popmintchev}. However, the efficiency of the high harmonic generation process decreases when employing laser fields at longer wavelength as it drops with $\lambda^{-5.5}$ \cite{tate07, perez09, yavuz12} which can limit the detection. Nevertheless experimentally, Colosimo \textit{et al.} have succeeded in studying the wavelength scaling of the harmonic generation employing multi-cycle laser fields at 0.8\,$\mu m$, 1.3\,$\mu m$, 2\,$\mu m$ and 3.6\,$\mu m$ \cite{colosimo}. In their case the Keldysh parameter was varied from 1.3 to 0.3. Henceforth, they could resolve the transition from multi-photon behavior to the classical limit at the longest wavelength. Besides experimental work, the theory was adapted to the wavelength region up to 5\,$\mu m$ to match the experimental data for high harmonic generation \cite{perez09}. \\

To access more complex dynamics under the non-perturbative regime Ghimire \textit{et al.} \cite{ghimire} demonstrated the generation of high harmonics in periodic solid structures using multi-cycle MIR laser sources. This experiment enabled access to intraband currents of these structures on the attosecond timescale. Transitions between bands of states within solids have already been described by Keldysh \cite{keldysh}. Thus, the transition rate of structure bands appears to be similar to that of the tunneling rate in atoms. This study has been extended further to cosine band structures for time-averaged transition rates \cite{gruzdev}.\\
Hawkins and Ivanov \cite{peter} have derived an analytic approximation for the transition rate in the low frequency limit to describe intraband sub-cycle electron dynamics. They showed that the transition depends on the band structure. This is due to electron acceleration after and during transition to the conduction band.
Therefore, to really access time-dependent sub-cycle electron dynamics in atoms, molecules or periodic structures, it is crucial to employ few-cycle NIR or MIR sources.
In this article we show the possibility to produce tunable few-cycle pulses via filamentation with central wavelengths tuned from 1.6 to 2\,$\mu m$. As a proof of principle we show how this source can be used to drive HHG experiments in atoms and molecules. This opens the potential to investigate sub-cycle dynamics of various targets in the non-perturbative regime. 

\section{Adiabatic versus non-adiabatic Keldysh parameter in the low frequency limit}

When $\gamma \ll 1$ which is the case for long wavelengths (low-frequency limit) and high intensities, the ionization rate for hydrogen atoms described by Keldysh \cite{keldysh} coincides with the well known tunnel ionization formula for static fields \cite{oppenheimer, landau}. In particular, the exponential dependence of the rate with respect to the field amplitude is present. When $\gamma \gg 1$, the formula reduces to a sum describing the ionization via the simultaneous absorption of several photons. In the tunneling case, i.e. $\gamma \ll 1$, the adiabatic approximation holds and the atomic structure can be reintroduced by means of an effective $n^*$, and the usual $\ell$ and $m$ quantum numbers in the exponential prefactor \cite{peremolov}. This approach has been further extended to any arbitrary ionic state by Ammosov \textit{et al.} \cite{adk} which is today known as the ADK theory. 

The results of these formulae are in excellent agreement with experimental measurements \cite{larochelle} accounting for cycle-average rates over one half-period of the alternating electric field, under the assumption of a sufficiently slow varying envelope between two consecutives cycles. With the possibility to produce and design few-cycle laser pulses, which is the frame of this article, this assumption is not valid anymore. Therefore, one has to use the non-adiabatic ionization rate formula developed by Yudin and Ivanov \cite{ivanov} for any arbitrary $\gamma$:
\begin{equation}
\Gamma(t)=N(t)\exp\left(-\dfrac{E_0^2f(t)^2}{\omega^3}\Phi\left(\gamma(t), \theta(t)\right)\right).
\label{NA_rate}
\end{equation}

\begin{figure}[h]
\includegraphics[angle=-90,width=\linewidth]{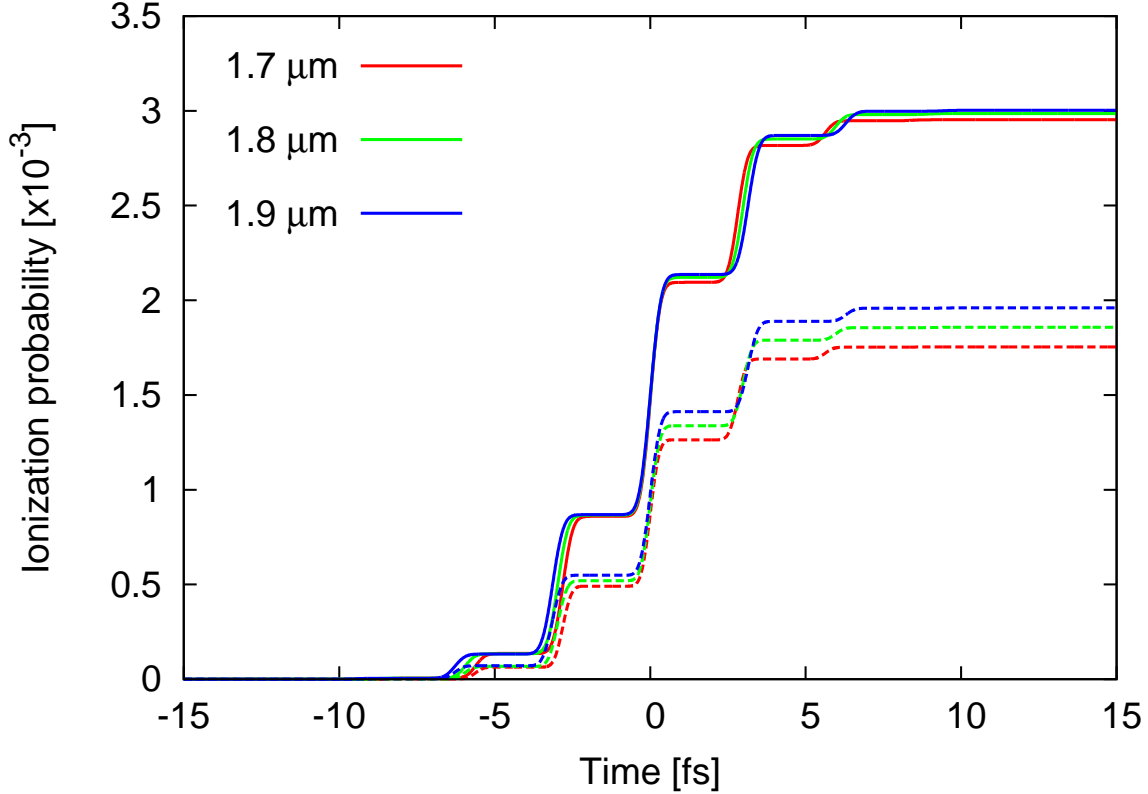}
\caption{Non-adiabatic ionization probabilities calculated in argon for 3-cycle Gaussian pulses, $\lambda$ = 1.7$\mu m$, 1.8$\mu m$ and 1.9 $\mu$m, $I$ = 10$^{14}$ W/cm$^2$ and $\phi_0$ = 0 (solid lines) compared to ADK ionization rates (dashed lines).}
\label{NA_IP}
\end{figure}

\noindent It explicitly takes into account the sub-cycle dynamics through the envelope $f(t)$ expressed in the time dependent Keldysh parameter $\gamma(t)=\gamma/f(t)$ and in $N(t)$ and through the instantaneous laser phase $\theta(t)=\omega t + \phi_0$ \cite{ivanov}. We applied this formula to the calculation of the ionization probability of argon by 3-cycle pulses generated at an intensity of 10$^{14}$ W/cm$^2$ and for 1.7$\mu m$, 1.8$\mu m$ and 1.9 $\mu$m wavelengths (cf. figure~\ref{NA_IP}). Once again, these long wavelengths ensure to be in 'better' tunneling conditions than the usual experiments at 0.8 $\mu$m and the 3-cycle pulse justifies the use of eq.~(\ref{NA_rate}).

\noindent Clearly, non-adiabatic ionization probabilities (solid lines) build up in a stepwise manner during the laser interaction but do not differ much at the end of the pulse for all considered wavelengths. By comparison, ADK ionization probabilities are lower by a factor of 1.5 to 1.7 and the probabilities are separated further when the interaction is over. This shows the strong influence of the sub-cycle dynamics on the ionization processes and the necessity to take them into account when a theoretical description is required in such conditions, i.e. for HHG experiments. It is worth mentioning that these two calculations asymptotically converge for $\omega\rightarrow 0$.

\section{Production of tunable NIR few-cycle pulses by filamentation}


Few-cycle pulses have been generated via various methods in the IR over the last decade. The two most common methods are hollow core fiber compression \cite{nisoli97} and filamentation \cite{guandalini,couairon07,hauri07,berge07}. These are used in a post-compression scheme after the amplifier system. \\
These techniques have been successfully adapted to the spectral region beyond the Ti:Sa wavelength of 0.8\,$\mu m$ and have been investigated experimentally as well as theoretically \cite{couairon07,hauri07,voronin11,berge08}.
In the 2\,$\mu m$ regime an optical parametric amplifier (OPA) scheme was used by Hauri \textit{et al.} generating 55\,fs, 330\,$\mu$J with carrier envelope stability \cite{hauri07}. By filamentation in a xenon cell they managed to compress the input pulse down to 17\,fs with 270\,$\mu$J pulse energy. 
Further, supercontinua spanning up to 3 octaves have been demonstrated by Kartoshov \textit{et al.} stretching from $\sim$\,0.35\,$\mu m$\,-\,5\,$\mu m$ \cite{kartashov12} employing a 80\,fs, 20\,Hz laser with a central wavelength of 3.9\,$\mu m$. M\"ucke \textit{et al.} \cite{mucke09} focused specifically on using 1.5\,$\mu m$ and managed to achieve a compression factor of 3\,-\,4 with pulses as short as 19\,fs and 1.5\,mJ pulse energy generated in an argon cell.\\
In parallel to these studies, the very well established hollow core fiber compression technique has been used at 1.8\,$\mu m$ to generate 11.5\,fs pulses at 400\,$\mu$J of pulse energy \cite{schmidt10}. These studies show how post-compression can be employed to produce few-cycle pulses in the NIR to MIR region. The next step is to show if conditions can be found to provide tunability of these sources.  
\\ 
A first study was performed using filamentation to generate tunable few-cycle pulses in the region of 1\,$\mu m$-\,2\,$\mu m$ \cite{trisorio} but it was not clearly shown if the pulse energy was sufficient to perform strong field interaction in the non-perturbative regime. 
We demonstrated the capability of filamentation to produce 1.6\,$\mu m$-\,2\,$\mu m$ tunable few-cycle pulses \cite{driever} and we report in this article how these pulses provide sufficient energy per pulse to produce non-perturbative HHG in Xe, Ar and $N_{2}$ molecules.



\section{Experiment}
\begin{figure}[h!]
\centering
\includegraphics[width=0.7\textwidth]{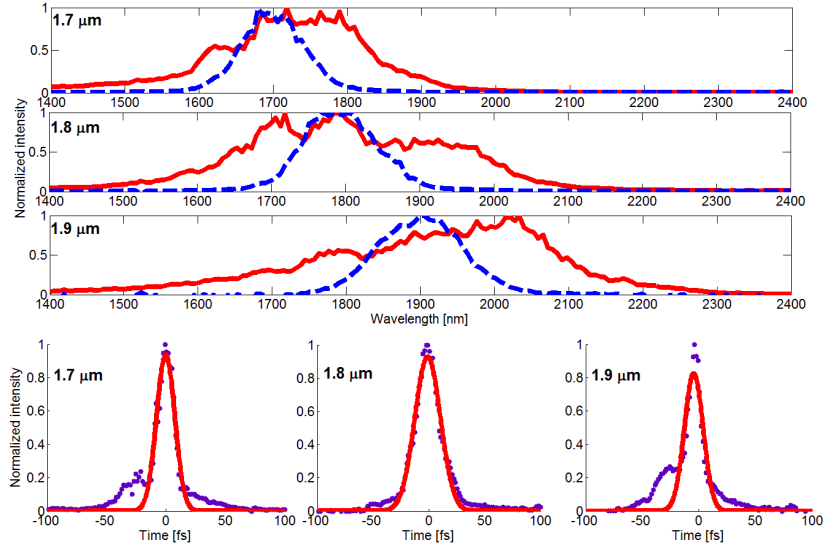} 
\caption{Upper panel: Input spectra (blue, dashed), spectra after filamentation (red, solid). Lower panel: auto-correlation for 1.7\,$\mu m$, 1.8\,$\mu m$ and 1.9\,$\mu m$. The purple dots are experimental data, the pulse width is given by the Gaussian fit. The fit gives 15\,fs for 1.7$\mu m$ and 1.9\,$\mu m$ whereas 1.8\,$\mu m$ reveals a pulse duration of 18\,fs. The Fourier limit of 1.7$\mu m$ and 1.9\,$\mu m$ was 12\,fs and the Fourier limit of 1.8\,$\mu m$ was 15\,fs.}
\label{fil}
\end{figure}

\begin{figure}[h!]
\centering
\includegraphics[width=0.4\textwidth]{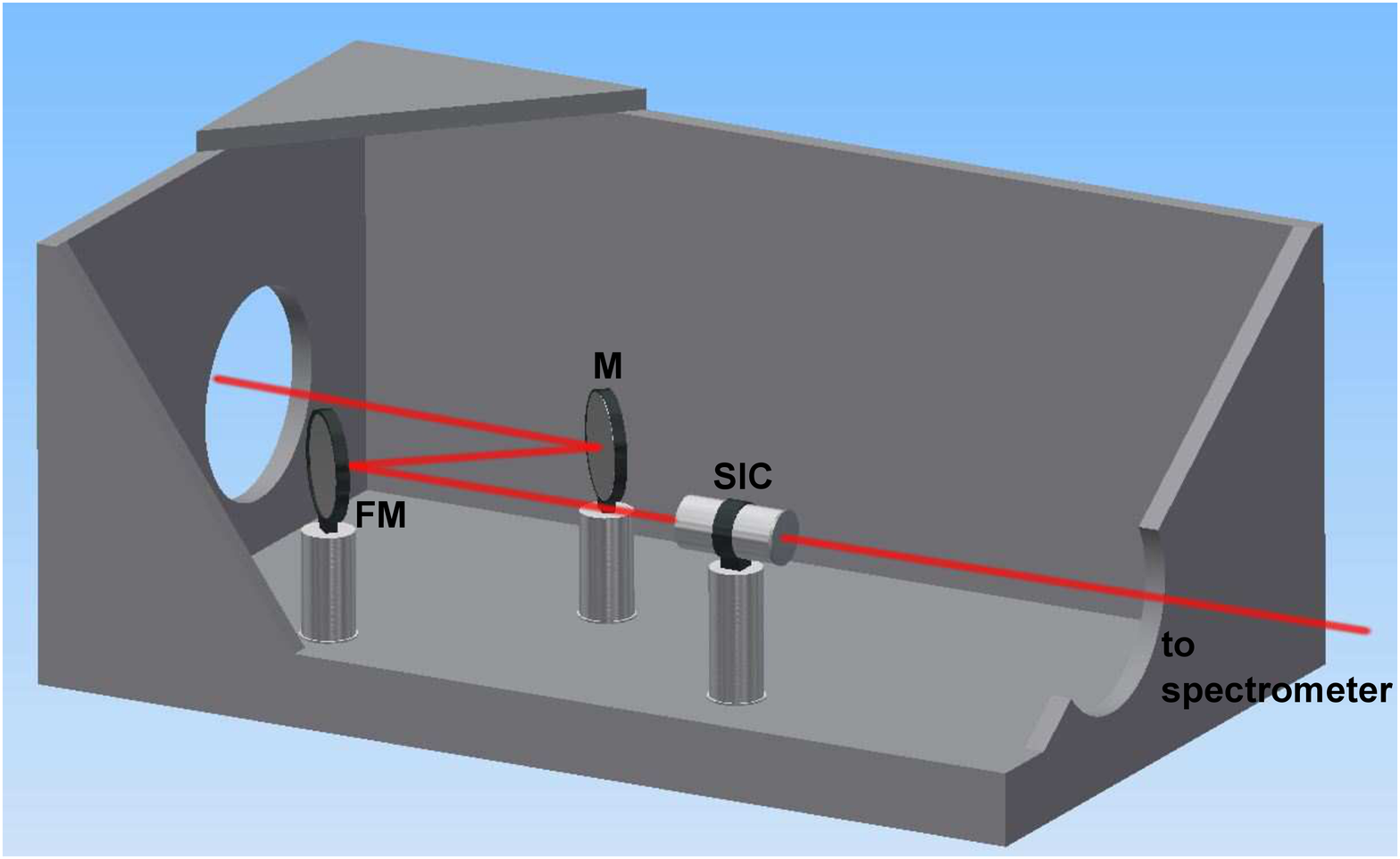} \includegraphics[width=0.42\textwidth]{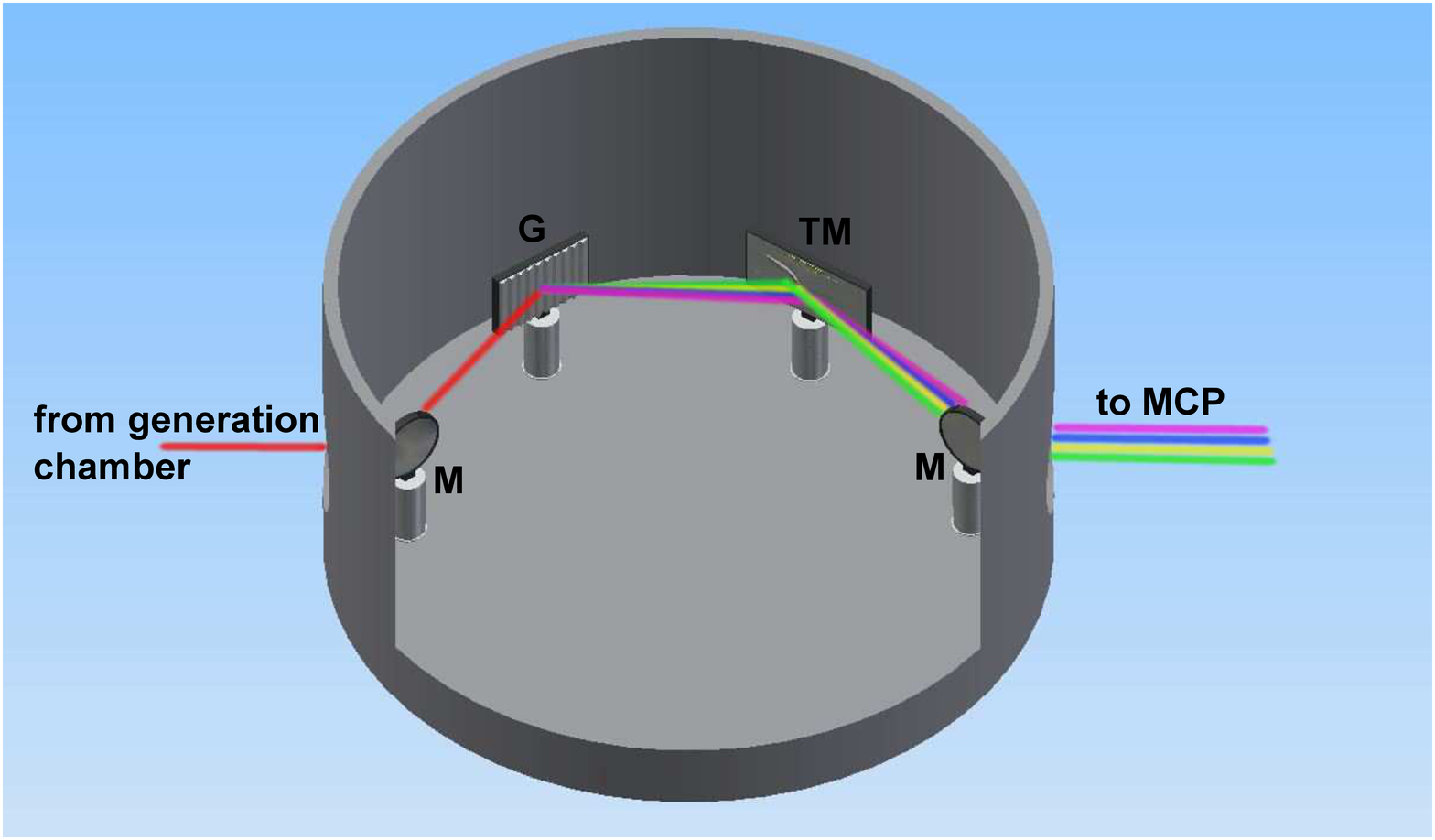}
\caption{Schematic of the experimental setup. After entering the generation chamber (left), the beam is redirected (mirror M) and focused (focusing mirror FM) to the interaction region in the semi-infinite cell (SIC). The beam then enters the XUV spectrometer chamber (right). The grating (G) disperses the harmonics and the toroidal mirror (TM) images the spectra onto the MCP.}
\label{setup}
\end{figure}

In order to generate the few-cycle pulses needed we employed an optical parametric amplifier (OPA) which generated 1.6\,$\mu m$\,-\,2\,$\mu m$ radiation with a pulse duration of 40\,fs. It was seeded by 0.8\,$\mu m$, 1\,kHz, 40\,fs, 5\,mJ pulses from a Ti:Sa amplifier. The generated NIR beam is produced with 0.7\,$\mu$J\,-\,0.9\,$\mu$J pulse energy and was loosely focused with a f=75\,cm concave mirror. The 1.2\,m cell was filled with 4\,bar of krypton. The cell was sealed with 1\,mm windows with an anti-reflection coating spanning 1.2\,$\mu m$\,-\,2.2$\mu m$. \\
In the generated filament, mainly self-phase modulation (SPM) contributes to the spectral broadening, but other effects like self-steepening can create a pedestal on the blue side. Further, ionization blue shift of the fundamental wavelength and high-order Kerr effects can also be involved.
We showed in a recent study  how tunable spectral broadening and compression from 1.6\,$\mu m$\,-\,2\,$\mu m$ can be produced \cite{driever}. Figure \ref{fil} shows an example of the tunable spectral broadening and compression obtained.
We managed to generate 3-cycle pulses for several central wavelengths of the OPA. The spectrum was broadened in the filamentation process by a factor of 2\,-\,3 to 250\,-\,300\,nm full width half maximum.\\
The accumulated positive group delay dispersion (GDD) was compensated by $\sim$\,4\,mm of fused silica. Since the fused silica has a negative GDD for these wavelengths it was robust and straight forward to compensate for the phase. For 1.7\,$\mu m$ and 1.9\,$\mu m$ we generated pulses of $\sim$\,15\,fs duration, whereas at 1.8\,$\mu m$ the duration was $\sim$\,18\,fs. The auto-correlation was very close to the Fourier limit of 12\,fs for 1.7\,$\mu m$ and 1.9\,$\mu m$ and 15\,fs for 1.8\,$\mu m$.  \\
We employed the 1.7\,$\mu m$ filament for our high harmonic generation (HHG) experiment after identifying the filament size \cite{amelle}. The centroid of the spectrum changed to 1.74\,$\mu m$ in the process. The size of the filament is defined as the region with the widest broadening and was estimated to $\sim$\,5\,mm.
Selecting this spatial region ensures that we only use the few-cycle pulses in our interaction region.\\
The high harmonics were generated subsequently in a semi-infinite cell (SIC) in vacuum, sealed with a 1\,mm fused silica window and an output stainless steel foil (150\,$\mu m$) at the exit. The 15\,fs pulses centered at 1.74\,$\mu m$ were focused by a f=30\,cm mirror (denoted FM) resulting in an interaction length of $\sim$\,30\,mm (cell size). The SIC output foil was drilled by the laser itself providing an additional differential pumping stage. The pressure can subsequently be used as an additional degree of freedom to enable efficient HHG. The beam size at the entrance and exit of the cell was estimated to be 260\,$\mu m$ and 65\,$\mu m$ respectively. Due to the fact that we focused on the output foil of the SIC, short and long trajectories for HHG can be phase matched \cite{salieres,QPI}. This however is mitigated by the long interaction region which results in a harmonic spectrum solely due to the short trajectories. The generated HHG is subsequently redirected to a 600\,lines/mm XUV grating (denoted G) at grazing incidence. The separated wavelengths are redirected with a toroidal mirror (denoted TM) onto the multi-channel plate (MCP) with phosphor screen and CCD camera. The experimental setup is shown in figure \ref{setup}.

\section{Results}
\begin{figure}[h!]
\centering
\includegraphics[width=\textwidth]{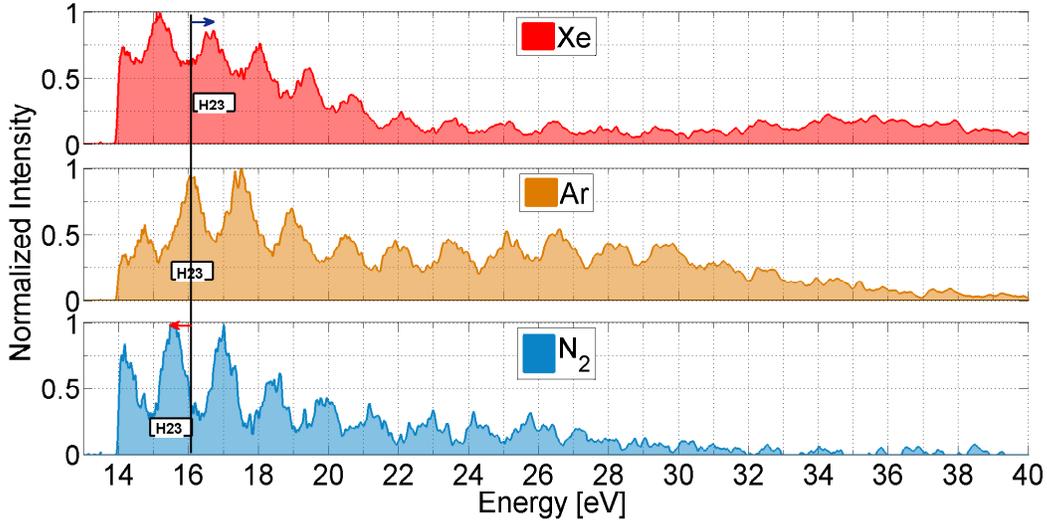} 
\caption{High harmonic spectra for xenon, argon and nitrogen for optimized generation conditions, top to bottom.}
\label{hhg}
\end{figure}

In figure \ref{hhg} from top to bottom the HHG spectra for xenon, argon and nitrogen are presented. The current observation window reaches from 14 to 50\,eV.
The xenon spectrum was acquired for a pressure of 41\,mbar and 180\,$\mu$J. This should lead to a cut-off energy of E$_{cut-off}$=(3.17U$_p$+1.3I$_p$) = 93.95\,eV (I$_p$(Xe)=12.13\,eV) at the focus which is at the end of the cell. Estimating the maximum photon energy from the intensity at the entrance of the cell we can expect a cut-off at 20.6\,eV.
The best spectrum for argon was obtained for 85\,mbar and 430\,$\mu$J potentially generating harmonics up to 200.9\,eV (I$_p$(Ar)=15.76\,eV) at the focus. Considering the intensity at the entrance of the cell we estimated  a cut-off energy of 32\,eV. The N$_2$ spectra on the other hand were generated at 105\,mbar and 320\,$\mu$J  pulse energy (E$_{max-cut-off}$= 154.8 eV, I$_p$(N$_2$)=15.58\,eV). The minimum cut-off energy should be about 28.4\,eV. The pressure ratio between xenon and argon $P_{argon}/P_{xenon} \sim 2.1 $ (respectively nitrogen molecules and argon $P_{argon}/P_{N_{2}} \sim 0.81$) is close to the square root of the inverse ratio of their masses $\sqrt{M_{xenon}/M_{argon}}=1.73$ (respectively $\sqrt{M_{N_{2}}/M_{argon}}=0.87$), so that the density of emitters for HHG in the volume of interaction is comparable.\\ 
In xenon (lower $I_{p}$ medium than argon) the high harmonic are blue shifted with respect to the argon high harmonics. As an example shown in figure \ref{hhg}, the $23^{rd}$ harmonic in xenon is blue shifted by half a photon energy at 1.74\,$\mu m$  with respect to the $23^{rd}$ in argon. This is due to the generation of plasma in xenon that modifies the fundamental pulse spectra towards lower wavelengths \cite{giammanco} as the saturation intensity of xenon ($8.2 \times 10^{13} W/cm^{2}$) is lower than in argon ($2.3 \times 10^{14} W/cm^{2}$) . The N${_2}$ HHG spectrum on the contrary is not blue shifted with respect to the argon (both have similar $I_{p}$ so similar saturation intensity) but it is red shifted. The $23^{rd}$ harmonic shown in figure \ref{hhg} is red shifted by one photon energy at 1.74\,$\mu m$ due to the Raman effect \cite{seideman}. The first Stokes shift due to a rotational effect from N${_2}$  (12\,$cm^{-1}$) induces a fundamental central wavelength shift of 1.5\,nm which is too small to compare to the spectral bandwidth of our few-cycle pulse. However over a long interaction region in the SIC, a cascade of first Stoke shift can occur resulting in a shift of multiple nm from the central wavelength (amplified Raman effect). 

\section{Conclusion and Outlook}

We managed to produce tunable few-cycle pulses in the NIR using filamentation. These NIR few-cycle pulses have sufficient energy per pulse to enable the generation of high order harmonics in the non-perturbative regime. The results presented using these pulses for HHG are very promising as they reveal effects of free electrons and Raman cascading encoded in the HHG spectra. This tunable source will enable the study of transitions between ionization regimes of atomic and molecular targets in gas or condensate phase. 

\section{Acknowledgment}
This work was supported by UK-EPSRC project EP/
J002348/1'CADAM',UK Royal Society project IE120539, Laserlab-Europe III CNRSCELIA001889, Inrex laserlab-europe III JRA, the region aquitaine (Nasa project) and the ANR attowave project. 
The authors acknowledge technical support from D. Descamps and F. Burgy from CELIA and P. Ruthven and A.
Gregory from Imperial College London.

\end{document}